# Fingerprints of a Local Supernova

Oliver Manuel[a] and Hilton Ratcliffe[b]

[a]*Nuclear Chemistry, University of Missouri, Rolla, MO 65401 USA*

[b]*Climate and Solar Science Institute, Cape Girardeau, MO 63701 USA*

**ABSTRACT.** The results of precise analysis of elements and isotopes in meteorites, comets, the Earth, the Moon, Mars, Jupiter, the solar wind, solar flares, and the solar photosphere since 1960 reveal the fingerprints of a local supernova (SN)—undiluted by interstellar material. Heterogeneous SN debris formed the planets. The Sun formed on the neutron (n) rich SN core. The ground-state masses of nuclei reveal repulsive n-n interactions that can trigger axial n-emission and a series of nuclear reactions that generate solar luminosity, the solar wind, and the measured flux of solar neutrinos. The location of the Sun's high-density core shifts relative to the solar surface as gravitational forces exerted by the major planets cause the Sun to experience abrupt acceleration and deceleration, like a yoyo on a string, in its orbit about the ever-changing centre-of-mass of the solar system. Solar cycles (surface magnetic activity, solar eruptions, and sunspots) and major climate changes arise from changes in the depth of the energetic SN core remnant in the interior of the Sun.

# INTRODUCTION

Fingerprints of a local supernova are not necessarily obscure and may be associated with such current issues as climate change, although that is not widely recognized in the debate over global warming. Repulsive interactions between neutrons—in the tiny nucleus that occupies a negligible fraction of the volume of individual atoms and stars—is the energy source that lights the Sun and warms planet Earth [1, 2], causes heavy nuclei to fission and massive stars to explode [3], and as will be discussed below, may drive the solar cycle of sunspots and the axial emission of material in the solar wind.

Fifty years ago, in October of 1957, Burbidge, Burbidge, Fowler and Hoyle [4, hereafter B2FH] published their classical paper on nuclear synthesis (nucleosynthesis) of elements heavier than hydrogen in stars. B2FH [4] were able to show that eight different types of nuclear reactions and a reasonable model of stellar evolution—from a hydrogen-rich first generation star to the final explosion of an evolved star as a supernova—could account for all of the isotopes of all elements heavier than hydrogen in the solar system today and also explain the correlation observed between nuclear properties and isotope abundances.

One year earlier, in 1956, two scientists who would later analyze meteorites and experimentally confirm important features of the B2FH scenario of element synthesis—especially r-products from the final supernova explosion—were actively pursuing related research projects. A young nuclear chemist who had studied nuclear fission at the University of Tokyo during Word War II, Dr. Kazuo Kuroda, correctly predicted that self-sustaining nuclear chain reactions (fission) occurred naturally in the Earth's geologic history [5, 6]. And a young physicist who had studied rare modes of nuclear decay at the University of Chicago, Dr. John H. Reynolds, developed the high sensitivity mass spectrometer [7] that would confirm—still preserved in the solar system today—decay products from short-lived isotopes and unmixed isotope anomalies from the stellar nuclear reactions that B2FH [4] would propose in 1957.

The B2FH paper [4] was the culmination of the nuclear era—a stage of science that had been feverishly pursued by competing nations during World War II and that held the entire world as hostage with the threat of mutual nuclear annihilation during the Cold War. The B2FH paper was published in October of 1957, and the Cold War competition ushered in the space age that same month when the Soviet Union successfully launched Sputnik I on October 4, 1957. The Cold War and the "space race" continued, and twelve years later precise isotope analysis on samples returned from the 1969 Apollo Mission to the Moon [8] provided the first hint of an unimaginable fingerprint—partially grasped at the time [9] but not fully recognized until 14 years later [10]—indicating that our Sun sorts atoms by mass and covers its surface with lightweight elements, but its interior consists almost entirely of even-Z elements of high nuclear stability from the deep interior of a supernova [10].

In 1960 Reynolds [11] reported the first hint of a supernova fingerprint in the material that formed the solar system—radiogenic $^{129}$Xe from the decay of extinct $^{129}$I ($t_{1/2}$ = 16 My) in a meteorite. Fowler, Greenstein and Hoyle [12] quickly noted that Reynolds' discovery was difficult to reconcile with the idea that the solar system formed from a typical interstellar cloud and suggested that $^{129}$I and other short-lived radioactive nuclei might have been produced locally, by less violent nuclear reactions than those that occur in a supernova. However, four years later Rowe and Kuroda [13] reported a less ambiguous supernova fingerprint in another meteorite—the decay products of extinct $^{244}$Pu ($t_{1/2}$ = 82 My). Unlike $^{129}$I, the radioactivity of $^{244}$Pu could not be explained by ordinary neutron capture on known nuclides in the solar system. The presence of $^{244}$Pu in meteorites required rapid neutron capture, the r-process described by B2FH [4], in a supernova at the birth of the solar system.

The most recent evidence for an energized, high-density supernova core lurking inside the Sun came from long-term geological studies of periodic changes in Earth's climate that are related to periodic changes in solar activity and to acceleration and deceleration of the Sun as it orbits the centre-of-mass of the solar system [14, 15].

The identification of each supernova fingerprint in the solar system initially seemed far-fetched and encountered stiff opposition. For example, this interpretation of the empirical link between primordial He in meteorites with excess $^{136}$Xe from the r-process of nucleosynthesis [16] was described in 1977 as *". . . too extreme to merit discussion"* [ref. 17, p. 209]. A decade later other critics stated that this idea is *". . . based on the use of an over-simplified model and a careless interpretation of the existing rare gas isotope data."* [ref. 18, p. 315].

Nevertheless, the empirical link of primordial He with excess $^{136}$Xe from the r-process of nucleosynthesis formed the basis for the successful 1983 prediction [10] that the Galileo Mission would find excess $^{136}$Xe from the r-process when the mass spectrometer on the Galileo probe entered Jupiter's He-rich atmosphere in 1996 [19].

Below is a summary of ten probable supernova fingerprints that have been recognized in the solar system since 1960, in approximately reverse chronological order—beginning with long-term, scholarly studies of periodicities in climate change and sunspots that might arise from a dense, energetic SN core inside the Sun and ending with isotope measurements that first revealed the decay products of short-lived radioactive nuclides from a supernova in meteorites and planets. These fingerprints are consistent with the scenario shown in Figure 1: The Sun exploded as a supernova five billion years (5 Gy) ago, giving birth to the planetary system [16]. After ejecting the material that now orbits it as planets, moons, comets, and asteroids, the Sun reformed as an iron-rich plasma diffuser around the neutron-rich supernova core [20].

The most obvious fingerprint of a local supernova—a striking similarity in the chemical layers of elements inside the Earth, across the solar system, and in the onion-skin model of a pre-supernova star—was not recognized until quantitative and precise experimental data showed that meteorites contain the decay products of short-lived nuclides and isotopic anomalies from stellar nucleosynthesis that are still linked today with elements from different layers of a heterogeneous, evolved star [16].

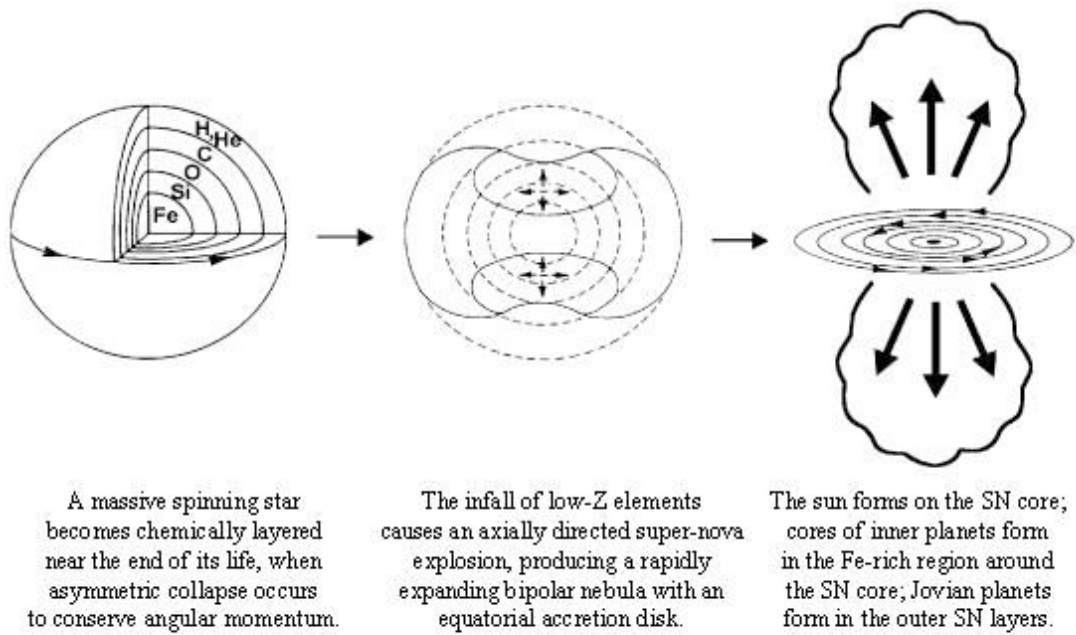

A massive spinning star becomes chemically layered near the end of its life, when asymmetric collapse occurs to conserve angular momentum.

The infall of low-Z elements causes an axially directed super-nova explosion, producing a rapidly expanding bipolar nebula with an equatorial accretion disk.

The sun forms on the SN core; cores of inner planets form in the Fe-rich region around the SN core; Jovian planets form in the outer SN layers.

**FIGURE 1.** This is a schematic drawing of the birth of the solar system, first revealed by the decay products of short-lived nuclides, isotopic anomalies from stellar nucleosynthesis reactions, and primordial He in meteorites that is tightly linked with excess Xe-136 from the r-process of nucleosynthesis [16].

## SUPERNOVA FINGERPRINTS

### I. Sunspots, Solar Cycles, Solar Inertial Motion, Climate Changes

Empirically observed links between sunspots, solar cycles, solar inertial motion, and changes in Earth's climate exposed the latest and most controversial fingerprint of a local supernova. As this paper goes to press, there is a great deal of debate and political wrangling over the possibility that the release of carbon dioxide into the atmosphere from the burning of fossil fuels has been the primary cause of global warming since the mid-20th century, about 1950.

Before this subject matter entered the political arena, many scholarly studies over much longer time periods [e.g., 14, 15, 21-24 and references therein] had shown that past climate changes were linked with solar activity and sunspots as the Sun moved in an irregular orbit about the centre-of-mass of the solar system.

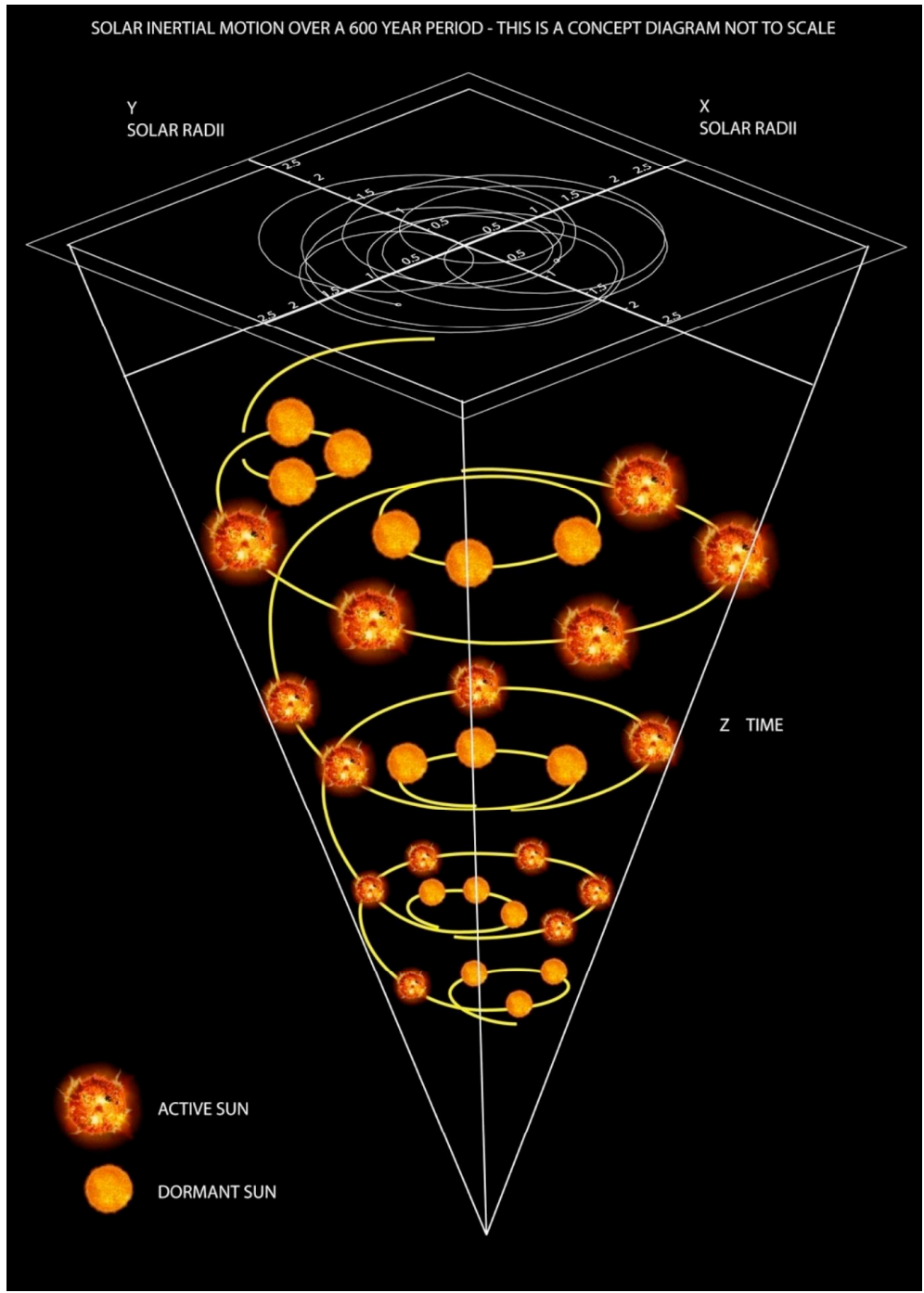

**FIGURE 2.** This schematic drawing illustrates sudden shifts in the solar inertial motion (SIM) as the Sun travels in an epitrochiod-shaped orbit about the centre-of-mass of the solar system. This 2006 drawing by Daniel Brunato, University of Canberra, was reproduced by permission from Richard Mackey [24].

For example, Alexander *et al.* [15] state that, "*Sunspot production is a direct function of the sun's galactic acceleration and deceleration . . .*" [ref. 15, p. 42] as the Sun moves like a yoyo on a string, in orbit about the constantly changing centre-of-mass (barycentre) of the solar system.

Figure 2 illustrates this uneven, jerky motion that the Sun experiences because of sudden acceleration and deceleration in its orbit about the barycentre of the solar system [24]. The gravitational force exerted by the planets, especially the more massive ones like Jupiter and Saturn, determines the position of the centre-of-mass of the solar system. The empirical link between planetary motions, sunspots, solar activity and climate change [14, 15, 21-24] contains valuable information on the internal structure of the Sun.

Figure 2 shows three complete orbits of the Sun, each of which takes about 179 years. Each solar orbit consists of about eight, 22-year solar cycles [24]. The total time span shown in Figure 2 is therefore three 179-year solar cycles [22], about 600 years.

The standard model of the Sun as a homogeneous ball of hydrogen remains very popular. That is perhaps why *"the mechanism for sunspot production as a result of galactic velocity changes in the sun has yet to be determined"* [15, p.42]. The observed link between sunspot production and changes in the sun's velocity would not be possible if the Sun were homogeneous [23, 24].

However, if the solar magnetic fields arise from iron-rich material in the deep interior of the Sun or from the compact neutron object at the solar core [25], then the location of this high-density material would likely shift relative to the visible solar surface as gravitational forces exerted by the major planets cause the Sun to experience abrupt acceleration and deceleration in its orbit about the ever-changing barycentre of the solar system [15].

We suggest that the empirical link between sunspots, solar cycles, solar inertial motion, and climate changes is the latest fingerprint of a local supernova—one that left a high abundance of iron and an extremely dense neutron core that moves about the solar interior as the Sun orbits the centre-of-mass of the solar system [14, 15, 21-24].

## II. Bipolar Outflows And Axially Directed Jets

Repulsive interactions between neutrons in the solar core generate solar luminosity, solar neutrinos, and the solar wind [e.g., 1, 2, 20]. If the solar cycle of sunspots is produced by changes in the location of that dense object in the interior of the Sun, as suggested above, then other puzzling features of the solar surface may arise from the emission of high energy neutrons in the solar core and be telltale signs of the collapsed supernova remnant there [3].

The anisotropic outflow of the solar wind, illustrated in Figure 3, is one such feature. This figure is from a 2005 NASA report on the primary mission results [http://ulysses.jpl.nasa.gov/science/mission_primary.html] of the Ulysses spacecraft. The top panel is an X-ray image of the Sun from the Soft X-ray Telescope on the Japanese Yohkoh spacecraft. The bottom panel shows the solar wind speed and density that the Ulysses spacecraft observed in measurements made at various latitudes, from near the South Pole to near the North Pole.

We suggest that the high-speed solar wind, 700-800 km/s, coming from the polar regions of the Sun is another fingerprint—faintly visible at the solar surface—of the compact object at the solar core. The solar-wind hydrogen is the decay product of neutrons emitted from the Sun's core [1, 2, 20], and the Spitzer telescope [26] has recently seen bipolar jets *"at a very early stage in the life of an embryonic sun-like star"* [http://www.spitzer.caltech.edu/Media/releases/ssc2007-19/release.shtml]. Bi-polar jets are common in compact stellar objects, and Adam Frank notes in a recent review, *"Bipolar outflows and highly collimated jets are nearly ubiquitous features associated with stellar mass loss"* [ref. 27, p. 241].

If asymmetry in the solar wind (Figure 3) begins by the same process that causes material to emanate axially from compact stellar objects, then this process may start with anisotropic emission from the solar core—as has been seen for $\alpha$-[28] and n-emission [29] from nuclei. Neutron-emission requires penetration of the gravitational barrier around a neutron star [1, 2, 20], just as $\alpha$-emission from a nucleus requires penetration of the Coulomb barrier. We suggest that neutrons are emitted axially from the solar core, perhaps for the same reason $\alpha$-particles are emitted in an axial direction from nuclei [28]: The shortest path through the barrier and thus the highest probability of barrier penetration may occur in an axial direction, because that is perpendicular to both the surface and to the spin direction. Higher angular momentum of particles at the equator reduces their probability of escape by sending the particle on a longer path through the barrier.

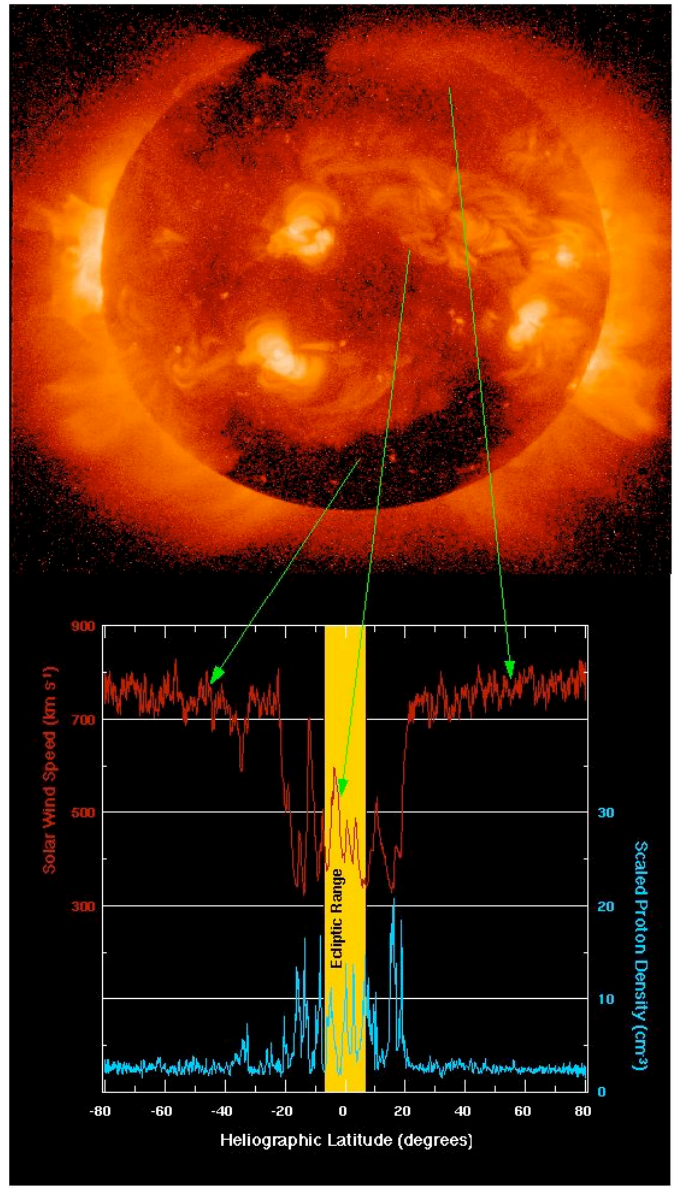

**FIGURE 3.** This is a summary of the primary findings from the Ulysses spacecraft, a combined NASA/ESA mission that orbited the Sun and measured properties of the solar wind as a function of solar latitude. This figure is from their report [http://ulysses.jpl.nasa.gov/science/mission_primary.html] dated 25 August 2005.

## III. Rigid, Iron-Rich Structures Below The Solar Photosphere

Figure 4 shows the remarkable fingerprint of a local supernova that was first noticed as rigid, iron-rich structures beneath the fluid photosphere in 2005 [30, 31], almost three decades after it was first suggested that a supernova gave birth to the solar system and the Sun formed on the collapsed SN core [32].

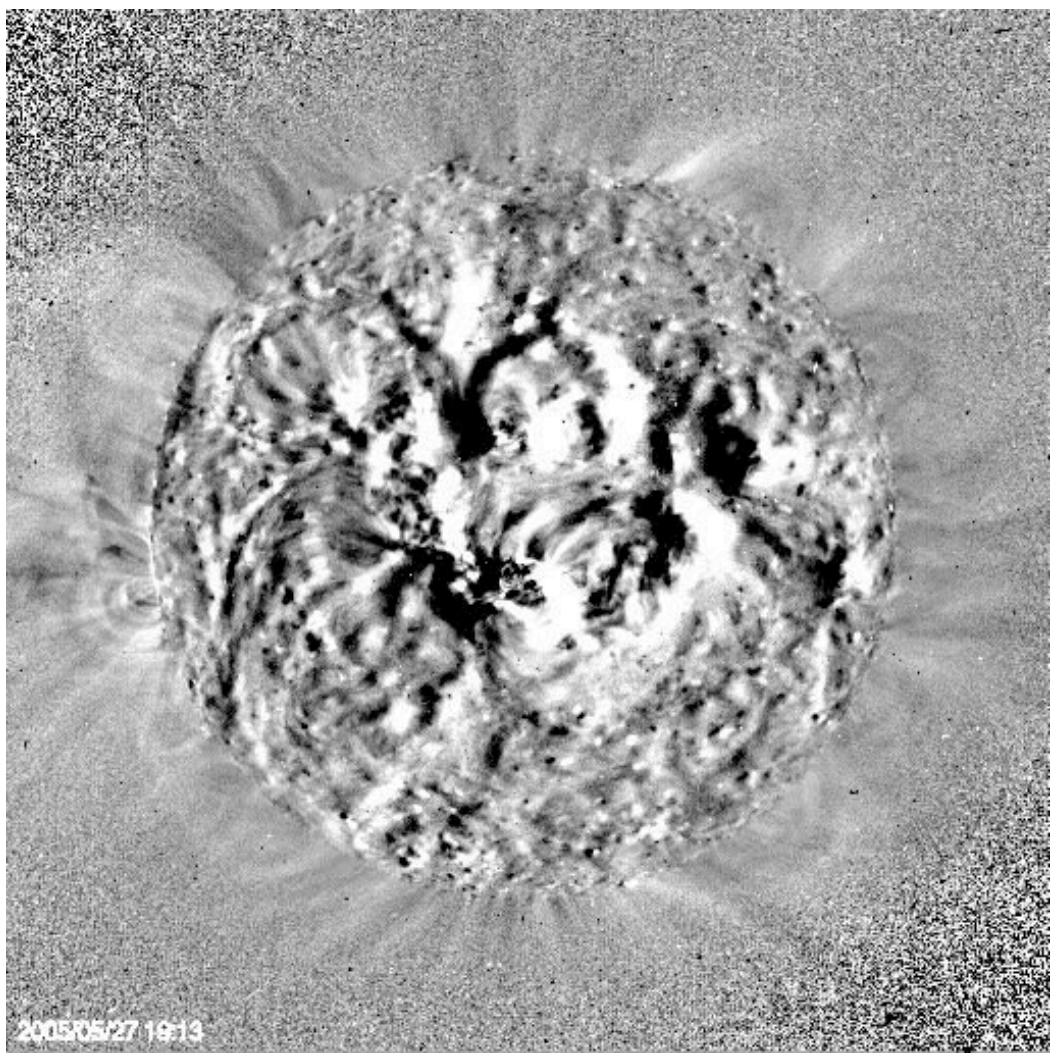

**FIGURE 4.** This is a "running difference" image of the Sun that was taken by the SOHO spacecraft on May 27th 2005 at 19:13, using a 195 Å filter to enhance light from emissions by Fe (IX) and Fe (X). These same rigid, iron-rich structures are visible for several days and even weeks later, rotating from left to right [30].

Michael Mozina, a software engineer in Mt. Shasta, CA, explained the impact of these solar images from the SOHO spacecraft on his opinion about the Sun [30]: *"While viewing images from SOHO's EIT program, I finally stumbled across the raw (unprocessed EIT images) marked "DIT" images that are stored in SOHO's daily archives. After downloading a number of these larger "DIT" (grey) files, including several "running difference" images, it became quite apparent that many of the finer details revealed in the raw EIT images are simply lost during the computer enhancement process that is used to create the more familiar EIT colorized images that are displayed on SOHO's website. That evening in April of 2005, all my beliefs about the sun changed"* [30, 31].

Mozina also noticed rigid, iron-rich structures in images that were taken with the TRACE satellite, using filters to enhance light emissions from iron ions [30].

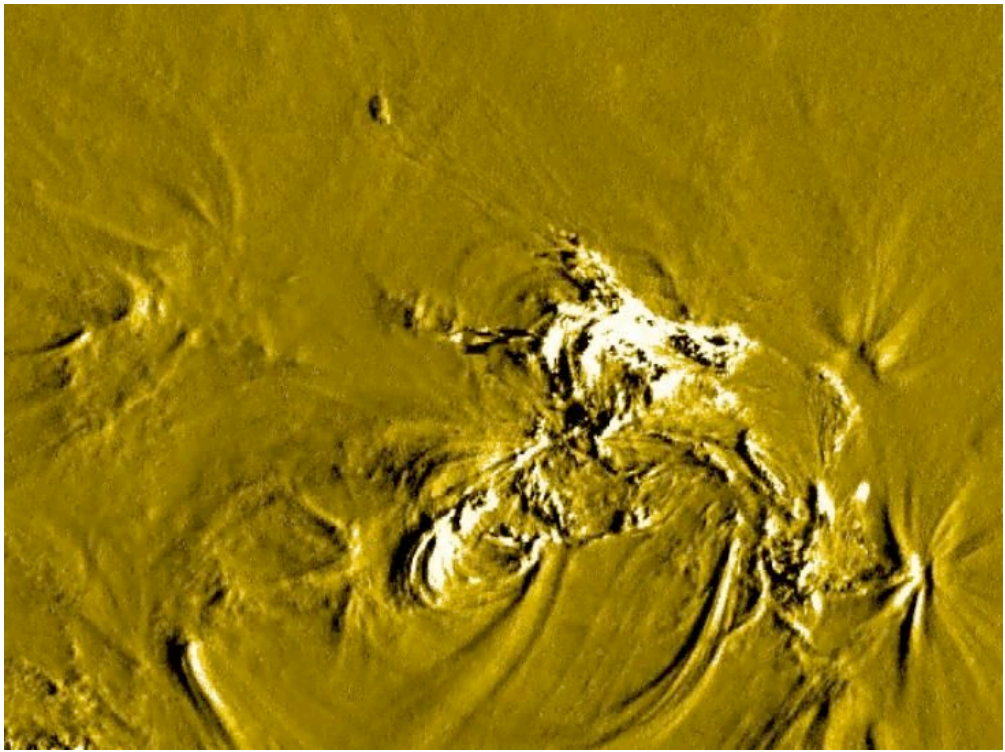

**FIGURE 5.** This is a "running difference" image of a small region of the Sun that the TRACE satellite recorded on 28 August 2000 using a 171 Å filter. This filter is specifically sensitive to light emissions from iron ions, Fe IX and FeX. The TRACE satellite later recorded an eruption and mass ejection from this same region. To view a video recording of the flare event, go to: http://trace.lmsal.com/POD/movies/T171_000828.avi

Figure 5 shows active region AR 9143, using the 171Å filter that is specifically sensitive to the iron ion (Fe IX/X) emissions. On 28 August 2000, an eruption and mass ejection occurred from this small region of the visible solar surface. A video recording of the mass ejection event can be viewed here: http://trace.lmsal.com/POD/movies/T171_000828.avi

## IV. Solar Luminosity, Solar Neutrinos, And Solar Hydrogen

As noted in the introduction, a systematic enrichment of lightweight isotopes in samples from the 1969 Apollo Mission to the Moon [8] provided an early hint [9] that the Sun sorts atoms by mass and selectively moves lightweight ones to its surface. Subsequent measurements showed a common mass fractionation pattern, extending from 3 amu to 136 amu (atomic mass units) [10]. Correcting the photosphere for mass fractionation yielded Fe, Ni, O, Si, S, Mg and Ca as the main constituents of the solar interior [10]—the same even-Z elements that constitute ≈ 99% of the material in ordinary meteorites [33] and in rocky planets close to the Sun. The probability of this remarkable agreement being a coincidence is essentially zero, $P < 2 \times 10^{-33}$ [34].

These seven elements all have high nuclear stability. Therefore, the source of solar luminosity, solar neutrinos and solar-wind hydrogen pouring from the surface of an iron-rich Sun remained a mystery for almost two decades, from 1983 until 2000.

That year five graduate students in an advanced nuclear science class at the University of Missouri-Rolla (Chem. 471)–Cynthia Bolon, Shelonda Finch, Daniel Ragland, Matthew Seelke and Bing Zhang–helped the first author of this paper construct a 3-D plot of reduced nuclear variables, M/A (mass per nucleon, or potential energy per nucleon) and Z/A (charge density, or charge per nucleon), showing each of the 3,000 known nuclides in the ground state [35]. The results, published on the cover of the book, "*Origin of Elements in the Solar System: Implications of Post 1957 Observations"* [27] and elsewhere [1, 2], are shown on the left side of Figure 6 as the "Cradle of the Nuclides."

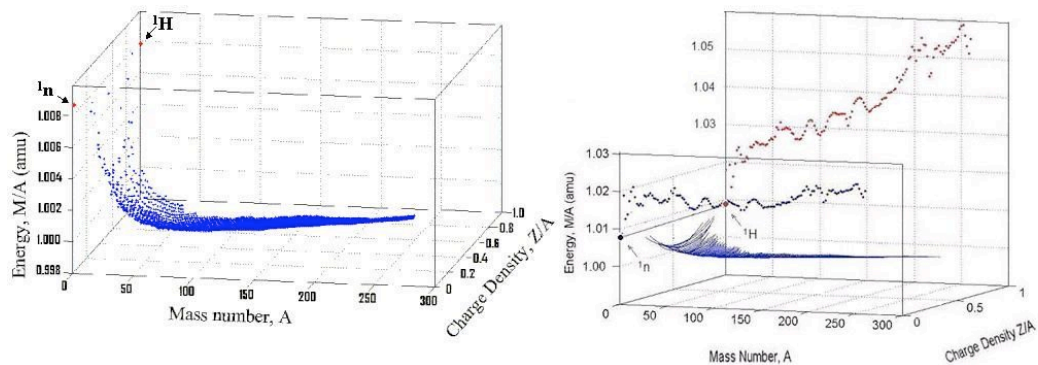

**FIGURE 6.** The "Cradle of the Nuclides" on the left shows the potential energy per nucleon for all stable and radioactive nuclides that were known in 2000 [34]. The more stable nuclides have lower values of M/A and occupy lower positions in the cradle. Nuclei that are radioactive or readily consumed by fusion or fission occupy higher positions. In the figure on the right, mass parabolas through data points at each value of A>1 intersect the front plane at {Z/A = 0, M/A = $(M/A)_{neutron}$ + ~10 MeV}.

Data points on the left side of Figure 6 represent each atomic mass in the ground state [35]. The right side of Figure 6 shows the mass parabolas defined by the data at each value of A>1 [35]. Intersections of mass parabolas with the front plane at Z/A = 0 show the potential energy per nucleon, M/A, for assemblages of pure neutrons. Intersections of mass parabolas with the back plane at Z/A = 1 show the potential energy per nucleon, M/A, for assemblages of pure protons. Repulsion between positive charges causes the value of M/A at Z/A = 1 to be larger than that at Z/A = 0 [1]. The intercepts at Z/A = 0 and Z/A = 1.0 display peaks and valleys at the same mass numbers because of tight and loose packing of nucleons, respectively [2].

Systematic properties of nuclei in Figure 6 reveal strong attractive interactions between unlike nucleons (n-p) and repulsive but symmetric interactions between like nucleons (n-n or p-p), that is further increased by Coulomb repulsion between positive charges on protons [1, 2]. Thus, empirical nuclear data indicate that a neutron is in an excited state in a neutron star, with about +10-22 MeV more energy than a free neutron [36]. On the other hand, theoretical studies suggest that neutron stars are "dead" nuclear matter, with each neutron having about -93 MeV less energy than a free neutron [37].

We prefer the conclusion based on empirical nuclear mass data [1, 2, 34, 36], and a recent review [38] agrees that useful information on neutron stars can be obtained by extrapolating atomic mass data out to *"homogeneous or infinite nuclear matter (INM)"* [ref. 38, p. 1042]. Thus solar luminosity, solar neutrinos and solar-wind hydrogen coming from the surface of an iron-rich object that formed on a collapsed SN remnant [16] are fingerprints of energetic neutrons in the solar core. These generate luminosity, neutrinos, and an outflow of 3 x $10^{43}$ $H^+$ per year in the solar wind by the following series of reactions [1, 2, 34, 36]:

1. Neutron emission from the solar core    Generates >57% of solar luminosity
    $<_0^1n> \rightarrow {_0^1}n + \sim 10\text{-}22$ MeV

2. Neutron decay    Generates < 5% of solar luminosity
    ${_0^1}n \rightarrow {_1^1}H^+ + e^- + \text{anti-}\nu + 0.78$ MeV

3. Fusion and upward migration of $H^+$    Generates <38% of solar luminosity
    $4\,{_1^1}H^+ + 2\,e^- \rightarrow {_2^4}He^{++} + 2\,\nu + 27$ MeV

4. Escape of excess $H^+$ in the solar wind    Generates 100% of solar-wind H
    3 x $10^{43}$ $H^+$/yr $\rightarrow$ Departs in solar wind

## V. The Iron-Rich Plasma Diffuser With A Hydrogen-Rich Veneer

Abundant lightweight elements at the solar surface, identified with large diamonds in Figure 7, were for several decades too familiar to be seen as strange fingerprints. Atomic abundances at the solar surface are ≈91% H, ≈9% He and ≈ 0.2% for all eighty-one heavier elements [39]. Hydrogen (H) is the lightest of all elements; He is the next-lightest one. Growing empirical evidence of severely mass fractionated isotopes [9, 11, 40-47] in the solar system pointed clearly to a solar site of the mass fractionation by 1983 [10], and this year two new papers report recent observational evidence that elements are also sorted by mass in other stars [48, 49].

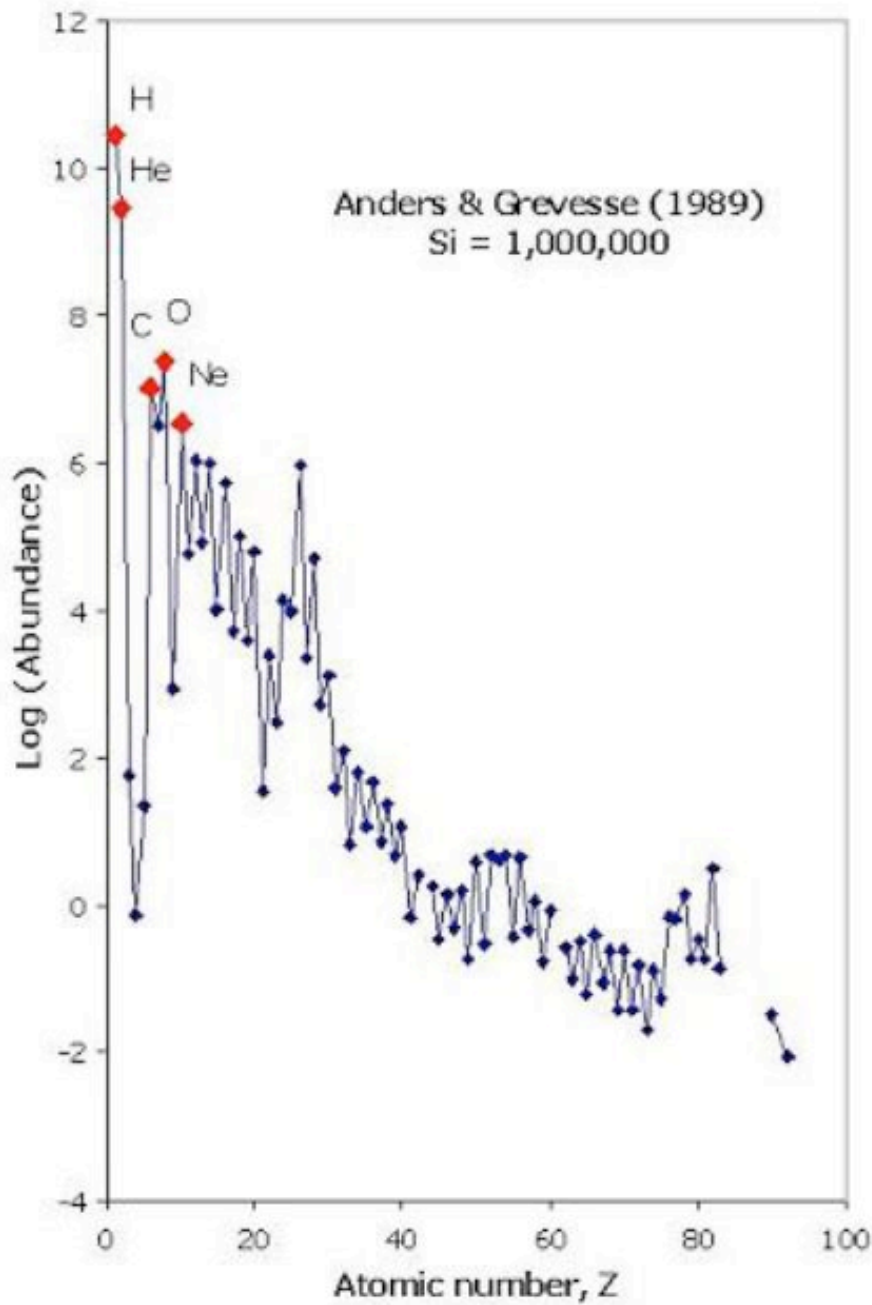

**FIGURE 7.** The abundance pattern of elements in the photosphere [39] generally declines with atomic (Z) number, except for a deficit of elements with unusually low nuclear stability (Li, Be and B) and an excess of those with unusually high nuclear stability (Fe and Ni).

Excess lightweight isotopes of He, Ne, Ar, Kr and Xe implanted in lunar soils by the solar wind revealed a common mass-dependent fractionation power law in 1983 [10], as shown in Figure 8. The abundance of a lighter isotope of mass (**L**) in the solar wind has been increased relative to that of a heavier isotope of mass (**H**) by a fractionation factor ($f$), where

$$\log (f) = 4.56 \log (H/L) \quad\quad (1)$$

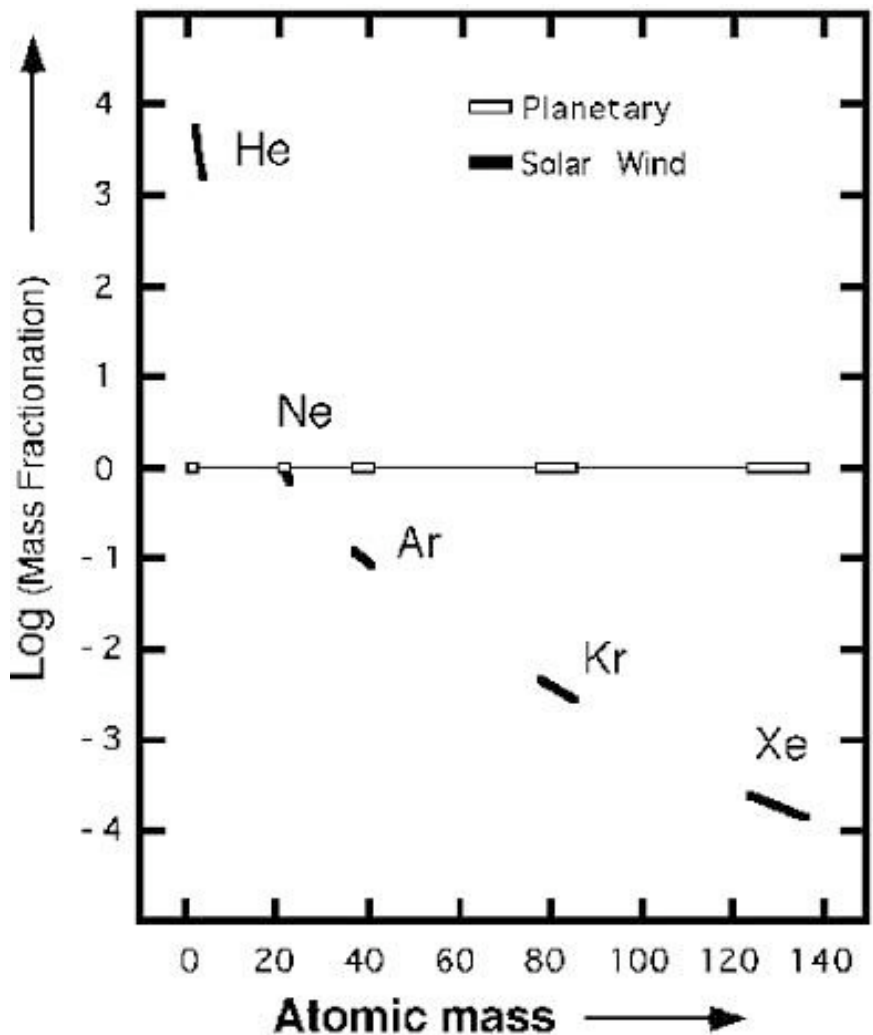

**FIGURE 8.** Isotope ratios of noble gases implanted in lunar soils exhibit a smooth, mass-dependent fractionation [10]. The percent enrichment per mass unit decreases as mass increases, from a value >200% per amu (atomic mass unit) for He to a value of 3.5% per amu across the nine stable isotopes of Xe. Mass fractionation is shown relative to $^{20}$Ne.

Figure 9 shows the abundance of elements in the bulk Sun, after correcting the abundance of elements in the solar photosphere [39] for the mass fractionation seen across the isotopes of noble gases in the solar wind [10]. The most abundant elements in the interior of the Sun are also the most abundant elements in ordinary meteorites— Fe, O, Ni, Si and S [33]. In 2005 it was shown that the fractionation pattern of s-products in the photosphere [50] also yields Fe, O, Ni, Si and S as the most abundant elements in the bulk Sun.

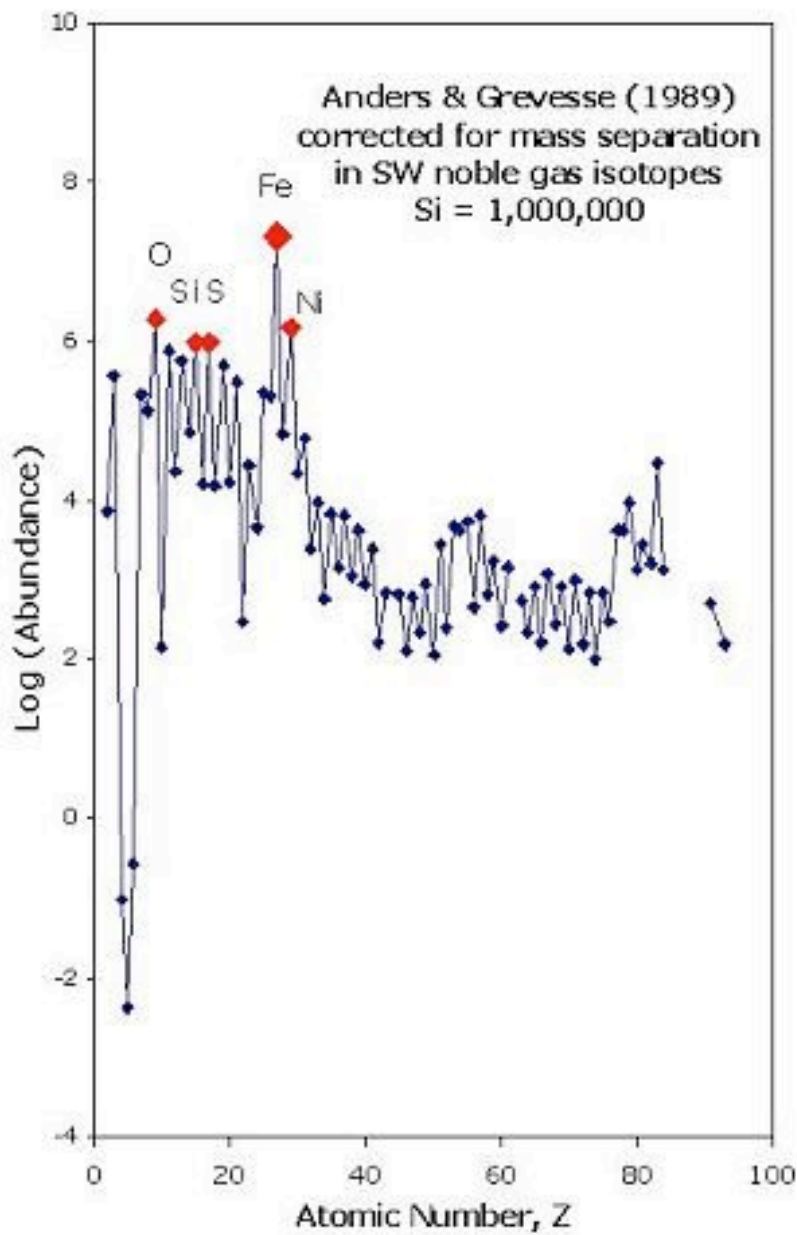

**FIGURE 9.** Large diamonds identify abundant elements inside in the interior of the Sun, calculated by correcting the abundance of elements in the photosphere [39] for the mass fractionation observed across the isotopes of elements in the solar wind [10].

# VI. Mirror Image Isotope Anomalies: Excesses And Deficits:

Reynolds encountered two SN fingerprints in xenon isotopes from meteorites in 1960: Radiogenic $^{129}$Xe from the *in situ* decay of SN-produced $^{129}$I [11] and isotope anomalies [51] from mass fractionation and stellar nuclear reactions. This was later called "strange xenon" [17] or **Xe-2** [49]—xenon that is enriched in $^{124}$Xe and $^{136}$Xe from the p- and r-processes of nucleosynthesis in a supernova [4].

A mirror image of "strange" xenon was discovered in silicon carbide (SiC) mineral separates of another meteorite in 1978 [53]. This mirror image component is enriched in the intermediate-mass xenon isotopes, $^{128,130}$Xe, which were made by the slow, s-process of nucleosynthesis [4], before the star reached the terminal supernova stage.

Precise isotope analysis of other heavy elements in meteorites and mineral separates of meteorites over the next fifteen years also indicated that "normal" isotope abundances of heavy elements may be a mixture of heterogeneous stellar debris that had still retained the signatures of distinct stellar nuclear reactions when refractory solids started to form in the solar system: Diamonds and graphite trapped heavy elements that were enriched in isotopes made by rapid r- and p-nucleosynthesis reactions in a supernova [4]; silicon carbide minerals trapped heavy elements that were enriched in the intermediate-mass isotopes that were made earlier by the slow s-process of nucleosynthesis [54-56].

Such nucleogenetic isotopic anomalies had been reported in six elements—krypton ($Z = 36$), tellurium ($Z = 52$), xenon ($Z = 54$), barium ($Z = 56$), neodynium ($Z = 60$) and samarium ($Z = 62$) by 1993, when Professor Begemann [56] brought attention to the striking "mirror-image" (+ and -) isotopic anomaly patterns in the isotopes of barium, neodynium, and samarium in mineral separates of meteorites.

Figure 10 is an illustration of the "mirror-image" isotopic anomaly patterns that Begemann noted in Ba, Nd and Sm from mineral inclusions of the Allende and Murchison meteorites [56]. Elements in inclusion EK-1-4-1 of the Allende meteorite (top) are enriched in isotopes that were made by rapid nuclear reactions in a supernova—the r- and p-processes of element synthesis [4]. These same elements in silicon carbide (SiC) inclusions of the Murchison meteorite (bottom) are enriched in intermediate mass isotopes that were made earlier by the s-process.

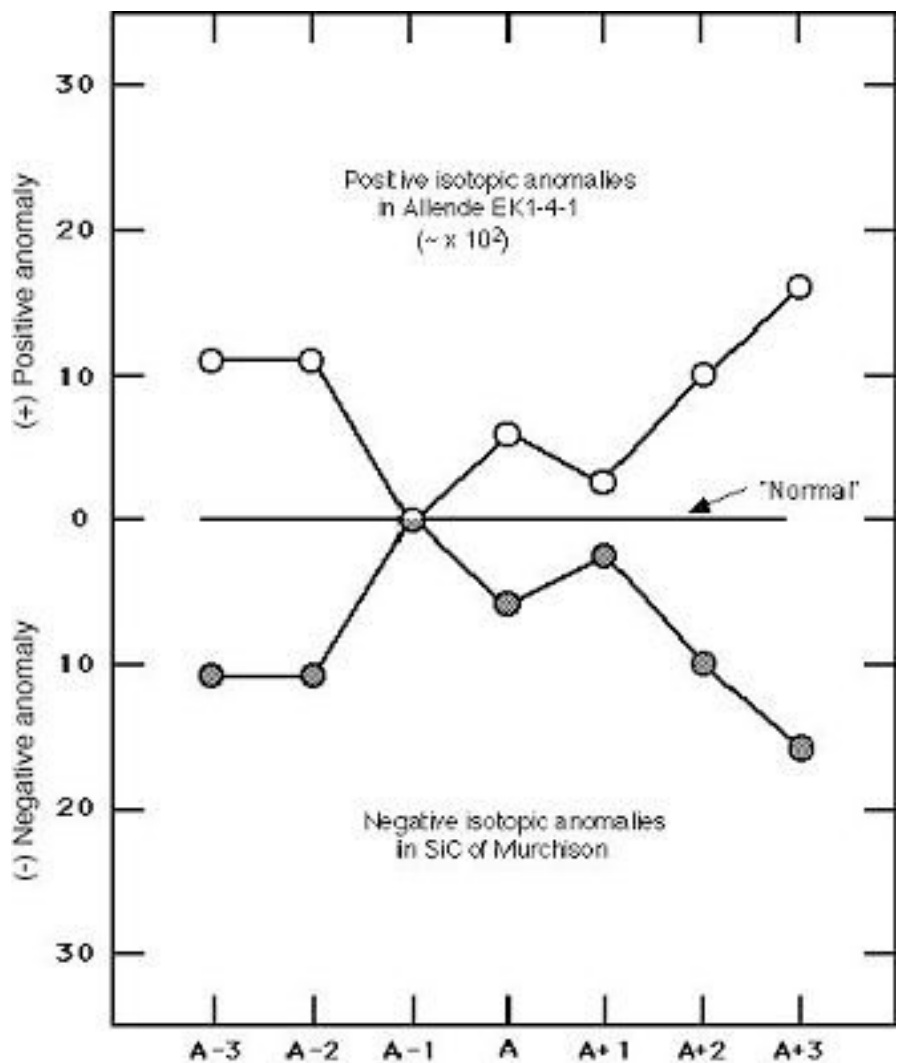

**FIGURE 10.** "Mirror-image" isotopic anomalies [56] observed in the isotopes of Ba, Nd and Sm from inclusion EK-1-4-1 of the Allende meteorite (top) and silicon carbide (SiC) inclusions of the Murchison meteorite (bottom). Isotope anomalies in these two meteorite inclusions are complementary, but the scale of s-products in the silicon carbide inclusion of the Murchison meteorite is about two orders-of-magnitude larger than that of the r- and p-products in inclusion EK 1-4-1 of the Allende meteorite [56].

The stellar debris that produced the isotopic anomaly patterns shown in Figure 10 was heterogeneous in both chemical and isotopic compositions. That is why the isotope record of nucleosynthesis by the r-, p- and s-processes still survives today in chemically distinct minerals and inclusions of meteorites; isotopes made by s-process were trapped in different minerals than isotopes made by the r- and p-processes.

Major elements formed silicon carbide (SiC) in a region of the heterogeneous stellar debris where s-products were prominent in the isotopes of heavy trace elements like Kr, Te, Xe, Ba, Nd, and Sm [53-56]. Carbon condensed into graphite or diamond (C) in a region of the heterogeneous stellar debris where r- and p-products were prominent in the isotopes of heavy trace elements like Kr, Te, Xe, Ba, Nd, and Sm [53-56].

## VII. Linked Chemical and Isotopic Heterogeneities in Planets

As mentioned above, the nucleogenetic isotopic anomaly pattern of a trace element may be linked with the presence of a specific major element. The link of $^{136}$Xe from the r-process with primordial He in meteorites [16-18] was the first example. Linked elemental and isotopic variations in elements of meteorites were pivotal to the search for variations in isotope abundances that are linked with major differences in the chemical compositions of planets and to the scenario shown in Figure 1 for the birth of the solar system from a supernova [16, 32].

Aside from detailed isotopic information on the material in planet Earth, only limited data are available on a few elements in two extra-terrestrial planets, Jupiter and Mars. Fortunately these two planets lie on opposite sides of the asteroid belt—the great chemical divide that separates the small, dense, rocky planets made mostly of Fe, O, Si, Ni, and S from the large, gaseous planets that are made mostly of the lightweight elements like H, He, C, and N.

Below is a list of the linked chemical and isotopic anomalies that have been confirmed to date in large objects in the solar system:

- The Galileo probe into the He-rich atmosphere of Jupiter in 1996 recorded excess $^{136}$Xe from the r-process of nucleosynthesis [19], like the excess $^{136}$Xe that that was trapped with "strange xenon" [57] and primordial helium in carbonaceous grains of meteorites at the birth of the solar system [16].

- The abundance pattern of xenon isotopes in Earth and Mars [58]— planets that are rich in iron (Fe) and sulfur (S)—matches the xenon isotope pattern seen in iron-sulfide (FeS) minerals of metallic [59] and stony [60] meteorites.

- Xenon in the Sun is mostly like that in Earth and Mars [58], but light isotopes are enriched at the solar surface by 3.5% per amu from solar mass fractionation. Only about 7-8% of the $^{136}$Xe in the Sun is the "strange xenon" that is seen in Jupiter and in carbonaceous grains of meteorites [61, 62].

- The level of mono-isotopic $^{16}$O in oxygen from the Earth is like that seen in differentiated meteorites, but less than that in carbonaceous meteorites [63].

- The latest measurements indicate that the level of mono-isotopic $^{16}$O (from He-burning) is lower in the Sun's oxygen than in the Earth or in carbonaceous meteorites [64], despite an earlier report that $^{16}$O is enriched in the Sun [65].

## VIII. Linked Chemical and Isotopic Heterogeneities in Meteorites

Figure 11 is a summary of a 1976 University of Chicago survey [63] on the level of mono-isotopic $^{16}$O in different classes of meteorites and their mineral separates. Oxygen-16 is a He-burning product—a nucleosynthesis fingerprint that distinguishes six categories of meteorites and mineral separates with distinctive chemical characteristics. This empirical evidence of linked chemical and isotopic variations in meteorites may be explained by a local supernova (Figure 1) [16] or perhaps by "mass-independent fractionation" at the solar system's birth [66, 67]. However, "mass-independent fractionation" [66, 67] does not explain the latest results that show a deficiency of $^{16}$O in the Sun [64].

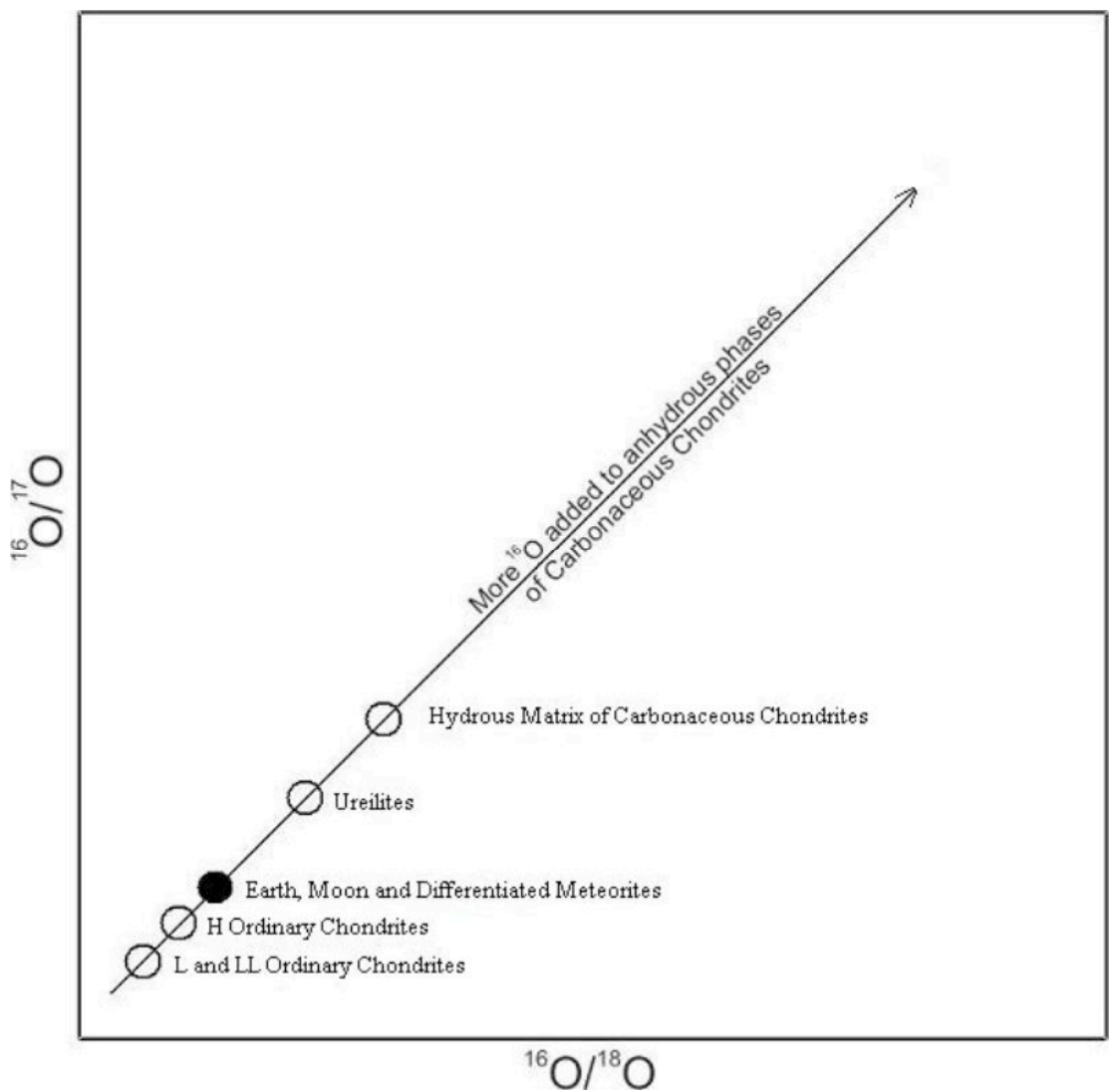

**Figure 11.** The abundance of $^{16}O$, a product of He-burning, is characteristic of six types of meteorites and mineral separates with other distinguishing chemical qualities [63].

Scientists at the University of Chicago also provided the first evidence that the abundance of another major element, helium, was linked with an isotopic anomaly from stellar nucleosynthesis when meteorites started to form [57]. Figure 12 shows the link of primordial He with excess $^{136}Xe$ from the r-process of nucleosynthesis in the Allende meteorite [32]. Subsequent analysis confirmed that the link of primordial helium with excess $^{136}Xe$ still remains as a SN fingerprint imprinted in the noble gases of meteorites [68, 69] and planets [19].

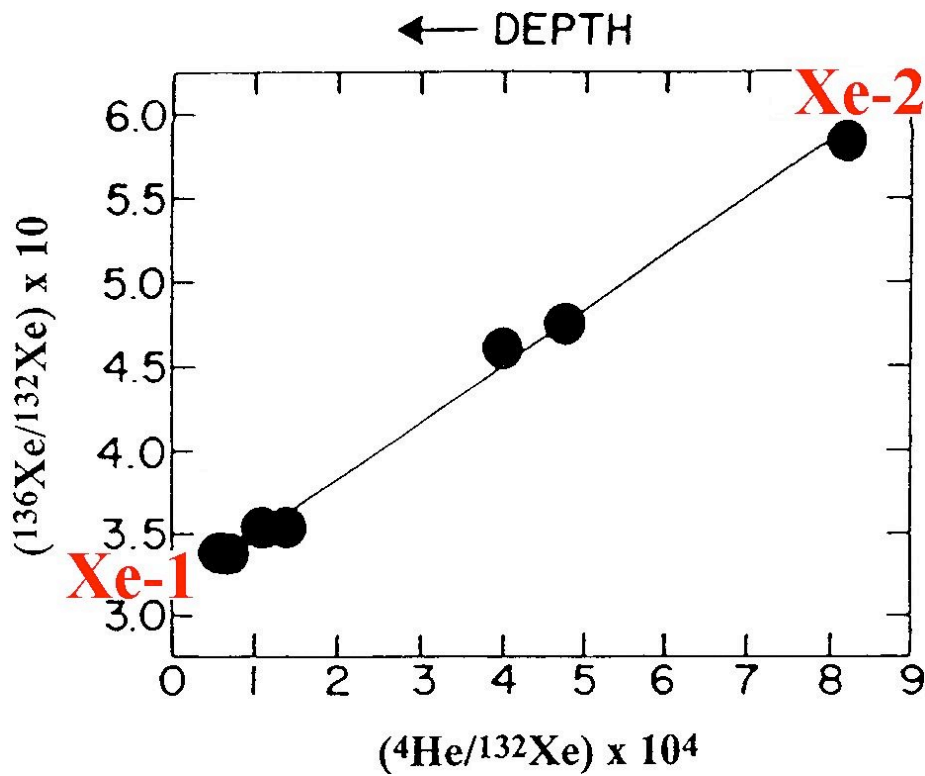

**Figure 12.** Analysis of the Allende meteorite in 1975 [57] first exposed a link [32] of primordial helium with excess $^{136}$Xe from the r-process. **Xe-1** is xenon of "normal" isotopic composition; **Xe-2** is "strange" xenon." The correlation suggests that $^{136}$Xe was made by the r-process in the outer layers of a supernova, where light elements like He were abundant. The arrow at the top of the figure indicates stellar depth [32].

Major elements that form the matrix of meteorite minerals are also linked with characteristic isotope signatures of nucleosynthesis in heavy elements:

- Carbonaceous (C) inclusions contain excess r- and p-products [54, 55].
- Silicon carbide (SiC) inclusions contain excess s-products and severely mass fractionated neon, Ne-E [47, 53-56].
- Iron sulfide (FeS) minerals contain little or no isotopic anomalies [59, 60].
- Iron nickel (Fe, Ni) contains small but significant isotopic anomalies [70].

## IX. Ubiquitous Isotope Anomalies From Nucleosynthesis

Anomalous isotope abundances from stellar nucleosynthesis were an unexpected SN fingerprint when first seen [52], but they now seem to be ubiquitous. They have now been identified in over twenty elements in meteorites [71, 72]—C, N, O, Ne, Mg, Si, Ar, K, Ca, Ti, Cr, Kr, Sr, Zr, Mo, Te, Xe, Ba, Nd, Sm, Dy, and Os. Meteorite-like isotope abundance patterns have also been recognized in Mars [58-60], in Jupiter [19], and in the Sun [61, 62, 64, 65]. Even the "normal" isotope abundances here on Earth appear in some cases to be mixtures of two anomalous components that were trapped in meteorites [56]. Only xenon isotope anomalies will be discussed in detail here.

Ubiquitous isotope anomalies across the solar system are likely fingerprints of the supernova (Figure 1) that gave birth to the solar system [16]. They are difficult to explain if 99.99% of the starting material of the solar system had "normal" isotopic composition and only 0.01% of the starting material had exotic nucleogenetic composition [73, 74].

The combined results of stellar nuclear reactions and mass fractionation were fortunately both noticeable in the unusual abundance pattern of xenon isotopes in meteorites when a nucleogenetic origin was first proposed in 1972 [52]. This is illustrated in Figure 13, where the solid, 45$^o$ line shows the correlation of r- and p-products of stellar nucleosynthesis. The approximately perpendicular dashed line shows the effects of physical mass fractionation in xenon isotopes [52].

Primordial isotope anomalies from nucleosynthesis were later found in many other elements, frequently embedded in an isotope spectrum that has also been altered by physical mass fractionation. Reasons for the association of nuclear and physical fractionation remained unknown and mysterious [45, 46] until mass fractionation was recognized as a common stellar process [10, 20]. For example, the term **FUN** isotope anomalies was coined in 1977 to identify the combined roles of **F**ractionation plus **U**nknown **N**uclear processes that generated correlated isotope anomalies of oxygen and magnesium in inclusions of the Allende meteorite [45, 46].

As shown by the dashed line in Figure 13, major xenon reservoirs in the inner part of the solar system—the Earth and the Sun—are related by mass fractionation of normal xenon (**Xe-1**). "Strange" xenon (**Xe-2**), enriched in r- and p-products, occurs in carbonaceous inclusions of meteorites and in Jupiter [19]. Bulk xenon in carbonaceous chondrites [51], **AVCC** Xe, is a mixture of **Xe-1** and **Xe-2**.

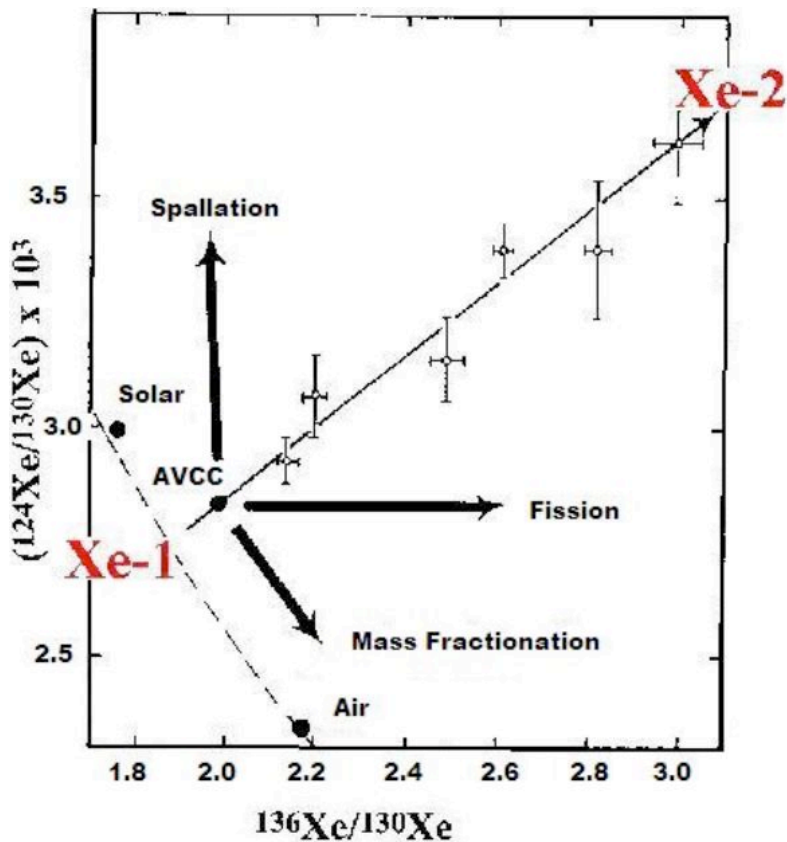

**FIGURE 13.** Xenon in meteorites is a mixture of normal xenon (**Xe-1**) with "strange" xenon (**Xe-2**) [49]. Mass fractionation (dashed line) of "normal" xenon (**Xe-1**) relates **solar** xenon with the xenon in **air**. "Strange" xenon (**Xe-2**) is normal xenon (**Xe-1**) plus extra $^{136}$Xe from the r-process and $^{124}$Xe from the p-process [4].

Figure 14 compares the "strange" abundance pattern of xenon isotopes in mineral separate 3CS4 of the Allende meteorite [57] with that in the Earth's atmosphere. Xenon in 3CS4 contains the smallest fractional amount of normal xenon, **Xe-1**. This measurement best represents the isotopic composition of **Xe-2**, which has also been referred to as **Xe-X** [52] and as "strange" xenon [57] in meteorites and in Jupiter [19].

Isotopic anomalies in xenon from the r- and p-process in Allende mineral separate 3CS4 (Figure 14) are mirrored as excesses of the intermediate mass xenon isotopes from the s-process of nucleosynthesis in silicon carbide grains from the Murchison meteorite [53]. The mass spectrometer that Reynolds built in 1956 [7] made possible the detection of more SN fingerprints in the nine isotopes of xenon than in any other element—as r-, p- and s-products of nucleosynthesis, as stellar mass fractionation, and as the decay of extinct $^{129}$I and $^{244}$Pu.

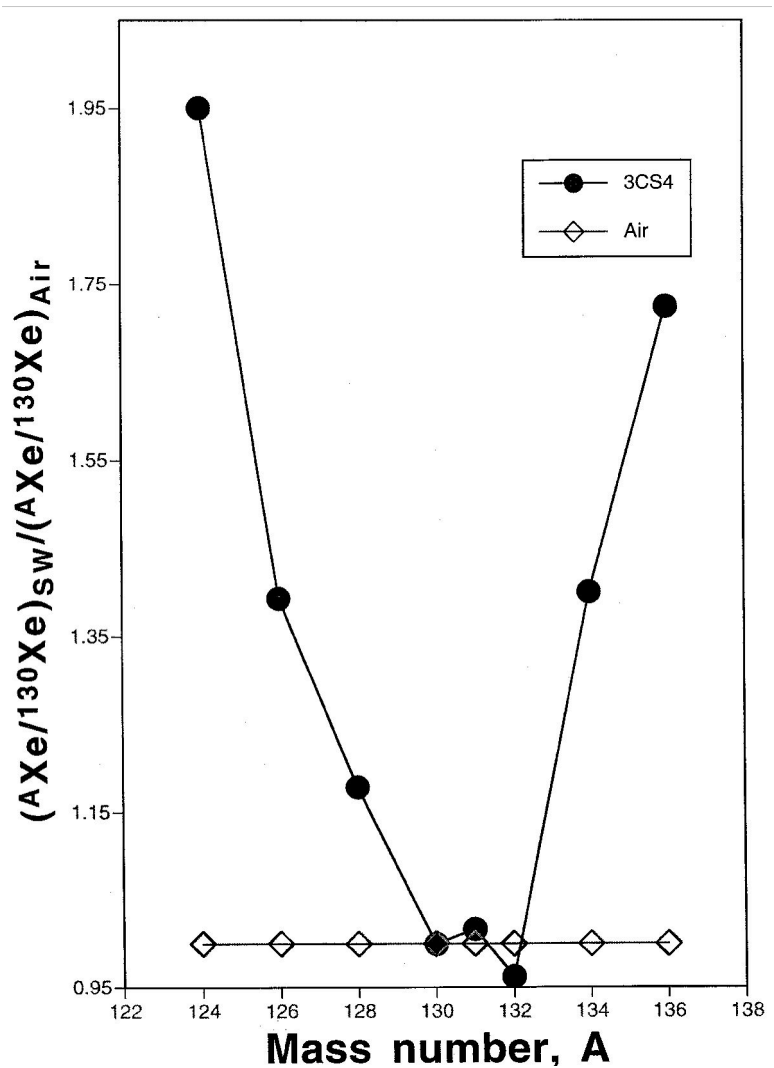

**FIGURE 14.** The p- and r-processes of nucleosynthesis [4] almost doubled the natural abundance of the lightest and the heaviest stable isotopes, $^{124}$Xe and $^{136}$Xe, of "strange" xenon trapped in Allende mineral separate 3CS4 [57]. This "strange" xenon was closely associated with the primordial helium incorporated into meteorites [68] and planets [19] at the birth of the solar system.

Anomalous isotope abundances in about twenty other elements recorded fingerprints of a supernova [71, 72]—a remarkably high fraction of those elements with the requisite three or more stable isotopes so that alterations from nucleosynthesis reactions, mass fractionation, and radioactive decay can be separated and identified.

Problems may still arise when mass fractionation is more severe than anticipated. The isotopic anomalies discovered in neon illustrate the problem. In the 1960s and 1970s, ten distinct isotopic forms of neon were identified in meteorites and labelled as Ne-A, Ne-B, Ne-C, Ne-D, Ne-E, etc. [53, 75, 76].

The mysterious origin of so many distinct forms of neon was finally solved in 1980, when it was shown that all of the isotopic forms of neon that had been reported in meteorites could be explained by severe mass fractionation and the addition of products of cosmic-ray-induced spallation reactions [47].

## X. Short-Lived Nuclides At The Birth Of The Solar System

Below is a chronological listing of short-lived nuclides whose decay products were reported in meteorites during the period of 1960 [11] to 2006 [13, 78-83].

| Short-lived Nuclide | Half-life | Reference |
|---|---|---|
| 1. Iodine-129 | 16 My. | J. H. Reynolds [11] |
| 2. Plutonium-244 | 82 My. | M. W. Rowe and P. K. Kuroda [13] |
| 3. Aluminum-26 | 0.7 My. | C. M. Gray and W. Compston [78] |
| 4. Palladium-107 | 6.5 My. | W. R. Kelly and G. J. Wasserburg [79] |
| 5. Manganese-53 | 3.7 My. | J. L. Birck and C. J. Allègre [80] |
| 6. Iron-60 | 1.5 My. | A. Shukolyukov and G. W. Lugmair [81] |
| 7. Calcium-41 | 0.1 My. | G. Srinivasan, *et al.* [82] |
| 8. Sodium-22 (???) | 2.6 y. | S. Amari, *et al.* [83] |

The 1960 discovery of radiogenic $^{129}$Xe in a meteorite [11] from the decay of extinct $^{129}$I ($t_{½}$ = 16 My) was the first hint of a supernova fingerprint in the solar system. In 1961 Fowler, Greenstein and Hoyle [12] suggested that local element synthesis might explain the unexpectedly high level of $^{129}$I radioactivity found in meteorites. The finding of $^{244}$Pu ($t_{½}$ = 82 My) decay products in a meteorite in 1965 [13] demonstrated the need for local element synthesis to include the r-process of nucleosynthesis [4].

All of the above short-lived nuclides may be products of a supernova [77], including the decay products of $^{129}$I and $^{244}$Pu that were found in the interior of the Earth in 1971 [84]. The decay products of shorter-lived nuclides in meteorites, and the absence of convincing evidence that any meteorite grains were exposed to cosmic rays during an interstellar journey to the solar system, favour the arguments that Fowler, Greenstein and Hoyle [12] made in 1961 for production of these short-lived radioactive nuclides by local element synthesis.

The supernova event shown in Figure 1 is compatible with the suggestion by Amari *et al.* [83] that $^{22}$Ne in graphite grains extracted from the Murchison meteorite *". . . is from the in situ decay of $^{22}$Na ($T_{1/2}$=2.6a) produced in the C-burning zone in pre-supernova stars",* and we hope that this research group has finally succeeded in finding excess $^{22}$Ne in a meteorite that is not accompanied by a proportionate excess of $^{21}$Ne from mass fractionation [47].

Kuroda and Myers [85] combined the U-Pb and $^{244}$Pu-$^{136}$Xe age dating chronometers to show that $^{244}$Pu in the solar system was produced in a supernova 5000 My ago. They also noticed [86] another convincing fingerprints of a local supernova when they compared physical properties and the amounts of radioactivity trapped in fallout grains from nuclear explosions with those in five special silicon carbide (SiC) grains that Amari *et al.* [87] recovered from the Murchison meteorite.

In both cases, the grains that formed first, grew larger and trapped more radioactivity than the grains that started to form later. The five special silicon carbide grains from the Murchison meteorite were physically larger and trapped more radioactive $^{26}$Al and more exotic isotopic anomalies than other silicon carbide grains. Mixing apparently occurred as short-lived nuclides decayed in the highly radioactive supernova debris that formed the solar system: Amari *et al.*

[87] state that the five special silicon carbide grains, "*. . . exhibit extremely exotic isotopic compositions, distinct from the majority of the SiC grains, . . .*".

## CONCLUSION

The Sun exploded as a supernova five billion years ago and ejected the material that now orbits it as planets, moons, asteroids, comets and other planetary rubble. Fingerprints of the supernova that gave birth to the solar system are visible in the planetary system as linked chemical and isotopic heterogeneities and as decay products of short-lived nuclides. Fingerprints of the supernova are visible in the Sun as overabundances of lightweight isotopes in the solar wind and overabundances of lightweight s-products in the photosphere, as solar emissions of energy, neutrinos, and hydrogen that match the products expected from a collapsed SN core—defined by systematic properties of the 3,000 known types of nuclei, as the axially directed solar wind, and as solar cycles (surface magnetic activity, solar eruptions, and sunspots) and changes in Earth's climate that are linked with gravitational forces that torque the Sun in its ever-changing orbit about the centre-of-mass of the solar system. The vibratory motion that planets induce in the star that they orbit may limit elemental segregation and bring enhanced metallicity closer to the surface of stars that host planets [88], like that ejected from the interior of the Sun by impulsive flares [89].

## ACKNOWLEDGMENTS

We are grateful to Professor D. Vincent Roach for reviewing this manuscript, to Richard Mackey for permission to reproduce the drawing by Daniel Brunato as Figure 2, to NASA and ESA for the summary of Ulysses spacecraft measurements shown in Figure 3 and the solar images shown in Figures 4 and 5 from the SOHO and TRACE programs, and to the Foundation for Chemical Research, Inc. (FCR) for permission to reproduce the other figures from reports to FCR, Inc. Students that were enrolled in Advanced Nuclear Chemistry (Chem. 471) in the spring semester of 2000 – Cynthia Bolon, Shelonda Finch, Daniel Ragland, Matthew Seelke and Bing Zhang – helped develop the "Cradle of the Nuclides" (Figure 6) that exposed repulsive interactions between neutrons as an important source of energy [1, 2, 35]. The moral support of former UMR Chancellors, Dr. Raymond L. Bisplinghoff (1975-1976) and Dr. Gary Thomas (2000-2005); the late Professors John H. Reynolds and Nobel Laureate Glenn T. Seaborg of UC-Berkeley; and former UMR Distinguished Curators' Professor and Chair of Chemistry, Professor Stig E. Friberg (1976-1979) is gratefully acknowledged. This paper is dedicated to the memory of the late Professor Paul Kazuo Kuroda (1917-2001) and his former student, Dr. Dwarka Das Sabu (1941-1991) for their contributions [9, 13, 16, 18, 31, 40, 46, 49, 58, 66, 82, 83] to the conclusions reached here and for their deep personal commitment to the basic precepts of science as a sacred path to truth. Finally, we remember with gratitude our colleague, the late Robert Bennett Blore, who quietly continued to take daily measurements on the Sun while we were arguing about theory.